\documentclass{emulateapj} 

\catcode`\"=\active\let"=\" 

\def\3{{\ss} }

\def\c12{{1\over 2}}

\def\plusplus{\raise 0.3ex\hbox{${\scriptstyle ++}$}{}}

\newcommand{\oversim}[2]{\protect{\mbox{\lower0.5ex\vbox{% 
   \baselineskip=0pt\lineskip=0.2ex 
   \ialign{$\mathsurround=0pt #1\hfil##\hfil$\crcr#2\crcr\sim\crcr}}}}}  
\newcommand{\simgreat}{\mbox{$\,\mathrel{\mathpalette\oversim>}\,$}} % >~ sign 
\newcommand{\simless} {\mbox{$\,\mathrel{\mathpalette\oversim<}\,$}} % <~ sign 
\begin{document} 
 
\title{The Cold Dark Matter Halos of Local Group Dwarf Spheroidals}  
%[CDM Halos of dSph galaxies] 
 
\author{Jorge Pe\~{n}arrubia\altaffilmark{1}, Alan W. McConnachie \& 
Julio F. Navarro\altaffilmark{2} }  
 
\affil{Department of Physics and Astronomy, University of Victoria, 
3800 Finnerty Rd., Victoria, BC, V8P 5C2, Canada} 
 
\altaffiltext{1}{Email: jorpega@uvic.ca} 
\altaffiltext{2}{Fellow of the Canadian Institute for Advanced Research} 
 
\begin{abstract} 
We examine the dynamics of stellar systems embedded within cold dark 
matter (CDM) halos in order to assess observational constraints on the 
dark matter content of Local Group dwarf spheroidals 
(dSphs). Approximating the stellar and dark components by King and NFW 
models, respectively, we identify the parameters of dark halos 
consistent with the kinematics and spatial distribution of stars in 
dSphs as well as with cosmological N-body simulations. Our analysis 
shows, in agreement with previous work, that the total mass within the 
luminous radius is reasonably well constrained and approximately 
independent of the luminosity of the dwarf, highlighting the poor 
correspondence between luminosity and halo mass at the extremely faint 
end of the luminosity function. This result implies that the average 
density of dark matter is substantially higher in physically small 
systems such as Draco and Sculptor than in larger systems such as 
Fornax. Because massive CDM halos are denser than low mass ones at all 
radii, these results imply that Draco formed in a halo 5 times more 
massive than Fornax's despite being roughly 70 times fainter.  Stellar 
velocity dispersion profiles ($\sigma_p(R)$) provide further 
constraints; in systems where data exist, $\sigma_p(R)$ remains flat 
almost to the nominal ``tidal'' radius, implying that stars are deeply 
embedded within their cold dark matter halos and are therefore quite 
resilient to tidal disruption. We estimate that halos would need to 
lose more than 90\% of their original mass before tides begin 
affecting the kinematics of stars, but even then the peak circular 
velocity of the dark halo, $V_{\rm max}$, would be barely affected. We 
estimate that $V_{\rm max}$ is about $3$ times higher than the central 
velocity dispersion of the stars, a result in agreement with previous 
estimates and that alleviates significantly the CDM ``substructure 
crisis''. We use these results to interpret the structural differences 
between the M31 and Milky Way (MW) dSph population and, in particular, 
the observation that M31 dwarfs are physically more extended by 
approximately a factor two than MW dwarfs of similar luminosity. Our 
modeling indicates that the difference in size should be reflected in 
their kinematics, and predicts that M31 dwarfs should have velocity 
dispersions up to a factor of $\sim 2$ higher than their MW 
counterparts. This is an eminently falsifiable prediction of 
CDM-motivated models of dSphs that may be verified with present 
observational capabilities. 
\end{abstract}

\section{Introduction}\label{sec:int} 
 
Ever since the first measurements of their velocity dispersion became 
available (Aaronson 1983, Aaronson \& Olszewski 1987), dwarf galaxies 
in the vicinity of the Milky Way have been a difficult puzzle to piece 
together in models of galaxy formation. Their relatively large size, 
low luminosity, and sizable velocity dispersion suggest the presence 
of large amounts of dark matter (Armandroff, Olszewski \& Pryor 1995, 
see the review of Mateo 1998; for more recent work consult Kleyna et 
al. 2005, Mu\~noz et al. 2005, 2006 and references therein). However, 
the absence of clear correlations between inferred dark matter content 
and the structural properties of the luminous component have led to 
arguments about the true physical nature of these objects and to 
various proposals to explain their large mass-to-light ratios.  
 
Velocity dispersions unduly affected by binary stars (Olszewski et al 
1996); line-of-sight alignment of unbound stars (Kroupa 1997), and 
remnants of stellar clusters severely disrupted by tides (Metz \& 
Kroupa 2007) have all been considered in the literature, but consensus 
now seems to have been reached. Dwarf spheroidal galaxies (dSphs, for 
short) are widely regarded as dark matter-dominated systems sampling 
the extreme low-mass end of the dark halo mass function. Nevertheless, 
a number of issues regarding the dark matter content of these elusive 
systems, as well as the spatial extent of their dark halos, remain 
largely unresolved. 
 
These issues were brought into focus when cosmological N-body 
simulations revealed the presence of substantial substructure in 
galaxy-sized cold dark matter (CDM) halos (Klypin et al. 1999, Moore 
et al. 1999). These simulations indicate that {\it hundreds} of 
self-bound cold dark matter halos massive enough (in principle) to 
harbor a dwarf galaxy are expected to populate the halo of the Milky 
Way, in sharp contrast with the mere {\it tens} of dwarf galaxies 
known to orbit the Galaxy. This realization rekindled interest in 
explaining the detailed correspondence between dark halos and luminous 
galaxies at the faint end of the luminosity function, an issue that 
had long been highlighted as a challenge for hierarchical galaxy 
formation models (White \& Rees 1978, Kauffmann, White \& Guiderdoni 
1993). 
 
The leading scenario for reconciling the discrepancy between 
``luminous'' and ``dark'' substructure in the Milky Way halo envisions 
dwarf galaxies as able to form only in halos above a certain mass 
threshold. The threshold is determined by the need to retain gas and 
to sustain continuing star formation despite the effects of feedback 
from evolving stars and the heating from photoionizing radiation 
(Efstathiou 1992, Bullock, Kravtsov \& Weinberg 2000, Somerville 2002, 
Benson et al. 2002). Adjusting the mass threshold appropriately, the 
scarcity of luminous dwarfs may be explained by the relatively few 
massive substructure halos that exceed the threshold (Stoehr et al 
2002, Hayashi et al 2003, Kazantzidis et al 2004). This is an 
appealing and elegant solution, and offers a relatively clean 
prediction that may be tested observationally: most dwarfs at the 
extreme faint end of the luminosity function should inhabit relatively 
massive halos, of mass comparable to that defined by the threshold. 
 
For example, given the large spread in luminosity of dwarfs in the 
Local Group{\footnote{We consider in this paper the ``traditional'' 
dwarfs brighter than $M_v \sim -8$ but recent discoveries based on the 
SDSS have uncovered the presence of many fainter dwarfs, extending the 
galaxy luminosity function to as faint as a few thousand $L_{\odot}$ 
(e.g. Irwin et al. 2007 and references therein)}}, and the expectation 
that they should all inhabit halos of similar mass, this proposal 
implies that there should be little correlation between the dark 
matter and luminous content of a dwarf. Massive dark halos would also 
be more affected by dynamical friction (Pe\~narrubia et al. 2002,  
Zentner \& Bullock 2003, Pe\~narrubia \& Benson 2005,  
Pe\~narrubia et al. 2006), leading 
to a bias in the spatial distribution of dSphs within the halo of the 
Milky Way relative to the bulk of substructure halos---an effect that 
may be potentially observable. 
 
Finally, their relatively massive halos would make these extremely 
faint galaxies more resilient to tidal disruption, suggesting that 
tidal tails and other obvious evidence of tidal stirring should be 
present only in systems that have lost most of their original mass by 
the action of tidal forces. In these models, the ``tidal'' radii 
attached by King-model fitting to the surface brightness profiles of 
dSphs, as well as the ``break'' radii identified in the outskirts of 
some dSphs, should reflect just the edge of the luminous component 
rather than a feature of dynamical significance (e.g. UMi, 
Mart\'inez-Delgado et al. 2001, Palma et al. 2003; Carina, Majewski et 
al. 2005). 
Establishing the present dark matter content of dwarf galaxies in the Local 
Group thus promises to be a fruitful enterprise. 
 
Validating (or refuting) these theoretical expectations through 
observation is, however, not straightforward. One reason is that few 
dwarfs have gas on circular orbits and, therefore, stars are the only 
viable dynamical tracer. The interpretation of the observations is 
thus complicated by degeneracies between orbital shapes, their radial 
dependence, and the overall mass profile (Wilkinson et al. 2002, 
Kazantzidis et al. 2004, Mashchenko et al. 2006, Kleyna et 2004, 
Tolstoy et al. 2004, Wilkinson et al. 2004, Mu\~noz et al. 2005, 2006, 
Wang et al. 2005, Walker et al. 2006a,b, Sohn et al. 2006). 
 
The second reason concerns the fact that these dynamical tracers lie, 
by definition, within the luminous radius of the dwarfs. Since dark 
halos are expected to be extended objects reaching far beyond the 
luminous confines of a galaxy, a certain amount of uncertain 
extrapolation appears inevitable. In this respect, mass-follows-light 
models (Richstone \& Tremaine 1986) only provide an approximate 
estimate of the amount of dark matter within the luminous radius that 
cannot be safely extrapolated to larger distances.  Finally, Galactic 
tides may affect dark matter and stars differently, especially if 
stars are, as expected, strongly segregated relative to the dark 
halo. Thus a dwarf may today inhabit a relatively low-mass halo (say, 
below the threshold mentioned above) even though it actually formed in 
a massive one that has since seen much of its mass stripped away by 
tides (Kravtsov, Gnedin \& Klypin 2004). 
 
This paper is the first in a series that attempt to address these 
issues by interpreting available observational data on Local Group 
dwarfs in the cosmological context defined by the leading paradigm of 
structure formation; the cold dark matter theory. We start by 
considering the constraints on CDM halos placed by the structural and 
dynamical properties of the dwarfs, and use these results to try and 
interpret the origin of structural differences in the population of 
dwarfs that orbit M31 and the Milky Way, respectively. A future paper 
on the subject will analyze in detail the effects of Galactic tides, 
both in the survival of the stellar components, as well as on the 
observational signatures imprinted by stripping (Pe\~narrubia, Navarro 
\& McConnachie in prep.). 
 
The paper is organized as follows. Section 2 introduces the relevant 
observations and the modeling procedure, whereas \S 3 applies the 
models to the Milky Way dwarfs. We apply these results to M31 dwarfs 
in \S 4, and conclude with a brief summary in \S5.

\section{Preliminaries}\label{sec:models} 
 
\begin{table} 
\caption{Observational properties of the Local Group Dwarf Spheroidals considered 
in this paper } 
\begin{tabular}{l l  l l l l l l } \hline \hline 
Name &  $\sigma_p(0)$ (km/s)&$R_c$ (kpc) & $R_t$ (kpc) & $D$ (kpc) & $l~(^\circ)$ & $b~(^\circ)$ & $M_v$ \\ \hline 
 Fornax & $10.5 \pm 2$     & 0.400   	&2.078      &   $138\pm 8$ & 237.1 & $-65.7$  &-13.0 \\ 
 Leo I  & $8.8 \pm 1$      & 0.169	&0.645      &   $250\pm 30$& 226.0 & $+49.1$  &-11.5 \\ 
 Sculpt & $6.6 \pm 1$      & 0.101	&1.329      &   $79\pm 4$  & 287.5 & $-83.2$  &-10.7 \\ 
 Leo II & $6.7 \pm 1$      & 0.162	&0.487      &   $205\pm 12$& 220.2 & $+67.2$  &-9.6 \\  
 Sextans& $6.6 \pm 1$      & 0.322	&3.100      &   $86\pm 4$  & 243.5 & $+42.3$  &-9.2 \\  
 Carina & $6.8 \pm 2$      & 0.177	&0.581      &   $101\pm 5$ & 260.1 & $-22.2$  &-8.6 \\   
 UMi    & $9.3 \pm 2$      & 0.196   	&0.628      &   $66\pm 3$  & 105.0 & $+44.8$  &-8.4 \\  
 Draco  & $9.5 \pm 2$      & 0.158	&0.498      &   $82\pm  6$ & 86.4 & $+34.7$   &-8.3 \\ \hline 
 And VII&$--$              & 0.450   	&4.300      &   $763\pm 35$& 109.5 & $-9.9 $  &-13.3 \\  
 And II &$(9.3 \pm 3)$       & 0.362   	&2.650      &   $652\pm 18$& 128.9 & $-29.2 $ &-12.6 \\  
% (And II) &         ---         & 0.990   	&4.200      &             &         &   & \\  
 And I  &$--$              & 0.580   	&2.300      &   $745\pm 24$& 121.7 & $-24.9 $ &-11.8 \\ 
 And VI &$--$              & 0.480   	&1.400      &   $783\pm 25$& 106.0 & $-36.3 $ &-11.5 \\ 
 Cetus  &$(17 \pm 2)$        & 0.290   	&7.100      &   $755\pm 23$& 101.5 & $-72.8 $ &-11.3 \\ 
 And III&$--$              & 0.290   	&1.500      &   $749\pm 24$& 119.3 & $-26.2 $ &-10.2 \\ 
 And V  &$--$              & 0.280   	&1.200      &   $774\pm 28$& 126.2 & $-15.1 $ &-9.6 \\ 
 And IX &($6.8$)-($12$)   & 0.296   	&1.300      &   $765\pm 24$& 123.2 & $-19.7 $ &-8.3 \\ 
\hline  
\newline 
\newline 
\end{tabular}\label{tab:obs} 
%\caption{The central velocity dispersion of the Milky Way dSphs was taken from Mateo (1998), whereas the structural parameters come from Irwin \& Hazidimitrious (1995). For the M31 dSphs the structural parameters were derived by McConnachie \& Irwin (2006) and the central velocity dispersions come from Lewis et al. (2006, Cetus); Chapman et al. (2005, And IX) and Cote et al. (1999, And II).}  
\end{table} 
 
\subsection{Summary of Observations} \label{ssec:obs} 
 
Table~\ref{tab:obs} lists the observational parameters of the Local 
Group dwarfs that we study here. We split the dwarfs into two groups, 
and list each in order of decreasing luminosity. The first set of 
objects are Milky Way (MW) dwarf spheroidals (dSphs), whereas the 
second set are all dSph type systems for which accurate core and tidal 
radii from King-model fitting are available. Most of the latter 
satellites orbit around M31, except for Cetus which is one of only two 
isolated dSphs in the Local Group. 
 
Table~\ref{tab:obs} lists the Galactic coordinates of each dwarf ($l$ 
and $b$), as well as its heliocentric distance, $D$. Distances are 
taken from McConnachie et al (2004, 2005) where available, otherwise 
from the compilation by Mateo (1998). Absolute magnitudes ($M_v$), 
core ($R_c$) and tidal ($R_t$) radii for the best-fitting King 
profiles to the surface brightness distribution are taken from Irwin 
\& Hatzidimitriou (1995) for the MW satellites and from McConnachie \& 
Irwin (2006) for Cetus and the M31 population, and are quoted after 
rescaling to the adopted distance. One exception is And IX, with core 
and tidal radii taken from Harbeck et al. (2005), distance from 
McConnachie et al. (2005), and the absolute magnitude from Zucker et 
al. (2004). The other is And II, the one dSph in the Local Group whose 
surface brightness profile shows evidence for the presence of more 
than one dynamical component (McConnachie \& Irwin 2006, McConnachie, 
Arimoto \& Irwin 2007) and where single King model fits are a poor 
description of the structure of the dwarf. In this case, we quote the 
And II core radius used by C{\^o}t{\'e} et al. (1999), since this is 
consistent with the traditional definition of core radius as the 
distance from the center where the surface brightness drops by a 
factor of two.

The central velocity dispersions ($\sigma_p[0]$) for the MW population 
in Table~\ref{tab:obs} are taken from the compilation by Mateo (1998) 
(his Table~7), although we note that since the publication of that 
review velocities for hundreds of stars in these dSphs have been 
measured with the aid of multi-object spectrographs. These data 
constrain the velocity dispersion profiles, $\sigma_p(R)$, in these 
systems, which are found to be approximately flat out to the nominal 
tidal radius (see, e.g., Fornax, Walker et al. 2006a, Battaglia et 
al. 2006; Leo~I, Koch et al. 2006; Sculptor, Tolstoy et al. 2004, 
Westfall et al. 2006; Leo~II, Sohn et al. 2006; Sextans, Walker et 
al. 2006b; Carina, Mu\~noz et al. 2006; Draco and Ursa Minor, 
Wilkinson et al. 2004, Mu\~noz et al. 2005). Note that the presence of 
several distinct components may affect the measured kinematics of a 
dwarf, depending on the spatial distribution of the tracers for which 
velocities are available (see, e.g., McConnachie, Pe\~{n}arrubia \& 
Navarro 2006, and references therein). We neglect here complications 
that arise from this issue, although we plan to address this in future 
work. 
 
Kinematic data for the M31 dwarf population is scarce and, 
comparatively, of poorer quality. For example, Chapman et al (2005) 
find that the velocity dispersion of And IX is $\sim 6.8 \pm 3$ km/s, 
but note that adding a single (possible member) star to their sample 
raises this estimate to $\sim 12$ km/s. Similarly, the velocity 
dispersion for And II (C{\^o}t{\'e} et al. 1999) is based on just 
seven stars and is thus subject to sizable uncertainty. The 
$\sigma_p(0)$ estimate for Cetus (Lewis et al 2006) is based on a 
larger sample of stars and it is thus more reliable. We present 
these data in Table~\ref{tab:obs} between parenthesis in order to 
emphasize that these data are not of the same quality as is available 
for the MW dSphs. We return to this issue in \S4. 
 
The past few years have seen a dramatic increase in the number of 
dwarf satellites discovered around M31 and the MW, in particular as a 
result of the completion of the Sloan Digital Sky Survey (Zucker et 
al. 2004, 2006a,b,c, Willman et al. 2005, Belokurov et al. 2006, 2007, 
Martin et al. 2006, Irwin et al. 2007). Many of these systems are 
morphologically not unlike the ones listed in Table~\ref{tab:obs}, but 
they are typically much fainter and of much lower surface brightness, 
which has precluded robust fits to their surface brightness profiles. 
 
The sample of Milky Way dwarf galaxies used here only includes (i) 
dwarf spheroidal galaxies (i.e., non-rotating systems with little or 
no detectable gas). These galaxies must have: (i) King profile fit 
parameters available from the literature and (ii) measured central 
velocity dispersion. Thus our sample excludes many of the recently 
discovered dSph galaxies in the Milky Way and M31. We have also 
excluded the Canis Major dSph (Martin et al. 2004)
because of its uncertain nature. A lively debate can be found in the literature, with arguments in favour of this system being a new galaxy (Mart\'inez-Delgado et al. 2005, Bellazzini et al. 2006) and against it (e.g. Momany et al. 2006, Moitinho et al. 2006). 
Sagittarius dSph was also removed from our sample because that is a clear case where tidal 
stripping has altered the luminous component as well as the dark 
matter halo. This particular system will be studied in detail in 
subsequent papers of this series, which address the effects of tides 
on the results presented here. 
 
\subsection{Modeling} \label{ssec:models} 
 
Our dwarf galaxies models assume the presence of two components in 
dynamical equilibrium: (i) a stellar component approximated by a King 
(1966) model, and (ii) a dark matter halo, which we approximate 
using an NFW profile (Navarro, Frenk \& White 1996, 1997). We assume 
that the dark matter dominates the dynamics of the system, and that 
stars may be regarded as massless tracers of the potential. 
 
\subsubsection{Luminous component} \label{sssec:kingmod} 
 
The density profile of a King model may be written as (King 1962) 
\begin{equation} 
\rho_\star(r)=\frac{K}{x^2}\bigg[\frac{\cos^{-1}(x)}{x}-\sqrt{1-x^2}\bigg], 
\label{eq:rhok} 
\end{equation} 
where 
\begin{equation} 
x\equiv\bigg[\frac{1+(r/r_k)^2}{1+(r/r_t)^2}\bigg]^{1/2}. 
\end{equation} 
Here $\rho_\star(r>r_t)=0$; $r_k$ is an inner radial scale; $r_t$ is 
the King ``tidal'' radius; and $K$ is an arbitrary normalizing 
constant. 
 
In the absence of rotation, the stellar kinematics is determined by 
the {\it total} gravitational potential, $\Phi(r)$, theough Jeans' 
equations. In particular, the radial velocity dispersion of stars, 
$\sigma_r$, is given by 
\begin{eqnarray} 
\sigma_r^2=\frac{1}{r^{2\beta}\rho_\star}\int_r^{r_t}r^{2\beta}\rho_\star\frac{d\Phi}{dr}dr, 
\label{eq:sr} 
\end{eqnarray} 
(Binney \& Tremaine 1987), where $\beta$ is the velocity 
anisotropy. We shall assume hereafter that the stellar velocity 
distribution is isotropic ($\beta=0$), which implies that the 
(observable) stellar line-of-sight velocity dispersion, $\sigma_p(R)$ 
is given by 
\begin{eqnarray} 
\sigma_p^2(R)=\frac{2}{\Sigma(R)}\int_R^{r_t}\frac{\rho_\star\sigma_r^2 r}{\sqrt{r^2-R^2}} dr, 
\label{eq:sp} 
\end{eqnarray} 
where $R$ is the projected radius and $\Sigma (R)$ is the projected 
stellar density, 
\begin{eqnarray} 
\Sigma (R)=2\int_R^{r_t}\frac{r\rho_\star}{\sqrt{r^2-R^2}}dr. 
\label{eq:Sigma} 
\end{eqnarray} 
We follow traditional convention and define the (projected) core 
radius, $R_c$, by the condition $\Sigma(R_c)=\Sigma(0)/2$. The core 
radius, defined this way, depends on both of the King model 
parameters, $r_k$ and $r_t$. For example, $R_c\simeq 0.73 \, r_k$ for 
$r_t/r_k=10$. 
 
\subsubsection{Dark matter component} \label{sssec:nfwmod} 
 
We assume that dark matter halos may be approximated by NFW profiles 
(Navarro et al 1996, 1997). The density profile may be written as 
\begin{eqnarray} 
\rho_{\rm NFW}=\frac{M_{\rm vir}}{4\pi r_s^3} \frac{(r/r_s)^{-1} (1+r/r_s)^{-2}} {[\ln(1+c)-c/(1+c)]}, 
\label{eq:rhoh} 
\end{eqnarray} 
where $M_{\rm vir}$ is the mass within the virial radius, $r_{\rm 
vir}$, $r_s$ is the scale radius and $c$ is the concentration 
($c\equiv r_{\rm vir}/{r_s}$). The virial radius is defined so that 
the mean overdensity relative to the critical density  is $\Delta_{\rm vir}$, 
\begin{equation} 
{M_{\rm vir} \over (4/3) \pi r_{\rm vir}^3}= \Delta_{\rm vir} \ \rho_{\rm crit}=\Delta_{\rm vir} {3 H(z)^2 \over 8 \pi  G}. 
\end{equation} 
We follow Bryan \& Norman (1998) and define the overdensity by  
\begin{equation} 
\Delta_{\rm vir}(z)=18\pi^2 + 82 \, f(z) - 39\, f(z)^2 
\label{eq:delv} 
\end{equation} 
where   
\begin{equation} 
f(z)=\frac{\Omega_0 (1+z)^3}{\Omega_0 (1+z)^3+\Omega_\Lambda}-1. 
\label{eq:omega} 
\end{equation} 
 
We assume throughout the paper a $\Lambda$CDM Universe by fixing the 
cosmological parameters to $\Omega_0=0.3$, $\Omega_\Lambda=0.7$, 
$h=0.7$, $n=1$, and $\sigma_8=0.9$, consistent with constraints from 
CMB measurements and galaxy clustering (see Spergel et al 2006 and 
references therein). 
 
Finally, we note that an NFW profile is fully determined by two 
characteristic parameters, such as the virial mass and the 
concentration. However, because the virial definitions adopted above 
depend on redshift, it is sometimes preferable to characterize an NFW 
halo by the location of the circular velocity peak, ($r_{\rm max}$, $V_{\rm max}$), where 
 
\begin{equation} 
V_{\rm max}\equiv V_{\rm NFW}(r_{\rm max})\simeq \bigg[\frac{GM_{\rm 
vir}}{2\, r_{s}}\frac{\ln (3)- 2/3}{\ln(1+c)-c/(1+c)}\bigg]^{1/2}, 
\label{eq:vmax} 
\end{equation} 
and  $r_{\rm max}\simeq 2 \, r_s$. 
 
Although the two characterizations are equivalent, we shall adopt the 
latter in this paper, since it is independent of redshift 
and its parameters are more easily compared with observation. 
 
\section{King models embedded in NFW halos}\label{sec:prof}

\begin{figure} 
\plotone{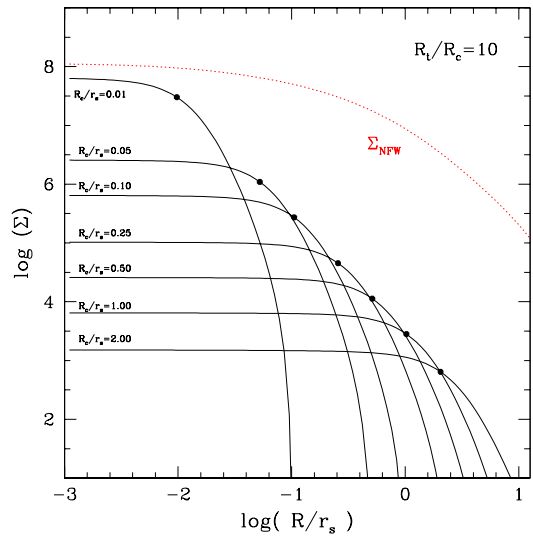} 
\caption{Stellar surface density profile of King models with the same 
``concentration'', $R_t/R_c=10$, and various degrees of spatial 
segregation relative to the dark matter. The segregation is measured 
by the ratio $R_c/r_s$, where $r_s$ is the scale radius of the NFW 
profile (shown by a dotted line in the figure) and $R_c$ is the 
stellar core radius, denoted with a dot in each profile. All radii 
have been scaled to the scale radius of the NFW profile. Units of 
surface density are arbitrary, since stars are assumed to contribute 
negligibly to the potential of the system.} 
\label{fig:dprof} 
\end{figure} 
 
\begin{figure} 
\plotone{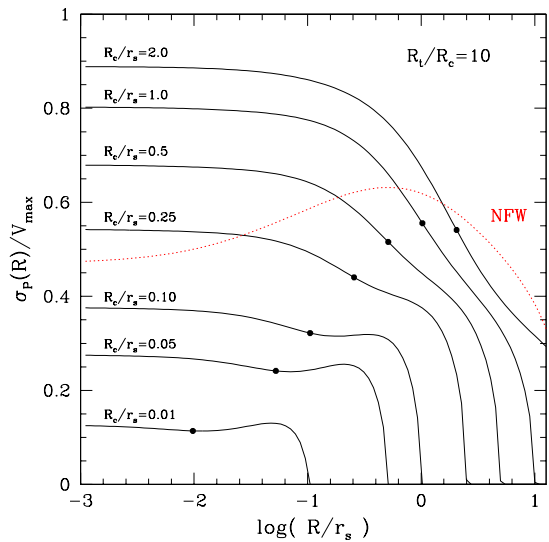} 
\caption{Stellar projected velocity dispersion profiles normalized to 
the halo maximum circular velocity for different values of 
$R_c/r_s$. The dotted line represents the projected velocity 
dispersion profile of the NFW halo. The solid circles indicate the 
core radius of the King model.  Note that both the central velocity 
dispersion of the stars, as well as the shape of the velocity 
dispersion profile, depend on the degree of spatial segregation 
between stars and dark matter (quantified here by the ratio 
$R_c/r_s$).} 
\label{fig:sprof} 
\end{figure}

\subsection{Spatial segregation and velocity dispersion} \label{ssec:seg} 
 
Stars that follow a King model and inhabit an NFW halo have their 
kinematics dictated principally by the spatial segregation of stars 
relative to dark matter, which may be quantified by the ratio of King 
``core'' radius to NFW scale radius, $R_c/r_s$. The more concentrated 
the stars are relative to the dark matter, the smaller the velocity of the 
stars will be relative to the characteristic circular velocity of the dark 
matter dominated potential.   
 
Fig.~\ref{fig:dprof} shows the projected 
density profile of King models  
as a function of $R_c/r_s$, and compares them to an NFW profile.  The radial units are scaled to the NFW scale radius, $r_s$, and the vertical units are arbitrary in this plot. The corresponding velocity dispersion profiles are shown in 
Fig.~\ref{fig:sprof}. Velocities have been scaled to the peak circular 
velocity of the NFW halo, $V_{\rm max}$, and show clearly the 
anticipated behaviour: the deeper the stars are embedded within the 
halo, the smaller the stellar velocities. For example, if $R_c\sim \, 0.1 
r_s$, the stellar central velocity dispersion, $\sigma_p(0)$, is only 
about 40\% of $V_{\rm max}$, but this value rises to 80\% for 
$R_c\sim r_s$. There is also a weak dependence on the tidal radius, 
but this is minor, as discussed below.

\begin{figure} 
\plotone{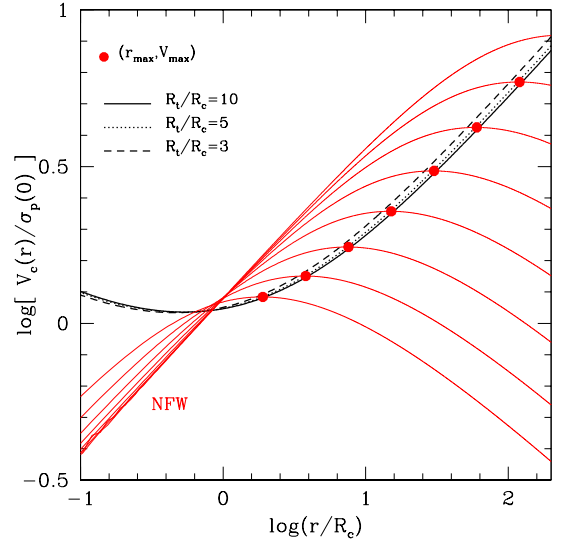} 
\caption{The King-NFW degeneracy. The thick lines show the halo peak 
circular velocity, $V_{\rm max}$, in units of the central velocity 
dispersion, and the radius of the peak, $r_{\rm max}$, in units of the 
King model core radius. Solid, dotted and dashed lines denote 
different King-model concentrations ($R_t/R_c$). Any NFW halo whose 
circular velocity peaks along this curve is consistent with the King 
model structure and kinematics. A few of these NFW models are shown 
for illustration by the thin curves.  Note that all these NFW models 
cross each other at approximately $R\simeq R_c$ and $V_c \simeq 1.2 \, 
\sigma_p(0)$. This implies that, given our assumptions, the mass 
within the core radius of the stellar component is robustly 
constrained to be $M(R_c)\sim 1.44 \, R_c \,\sigma_p^2(0)/G$.} 
\label{fig:s0rcrs} 
\end{figure}

This implies that a ``family'' of NFW halos is consistent with a King 
model of given $R_c$ and $\sigma_p(0)$ (see Strigari et al 2006 for a 
similar argument applied to the Fornax dSph). The more embedded we 
assume the King model to be within the NFW halo, the more massive 
(i.e., higher $V_{\rm max}$) the halo must be in order to explain a 
given $\sigma_p(0)$. This ``King-NFW degeneracy'' is illustrated in 
Figure~\ref{fig:s0rcrs}, where the thick lines show the correspondence 
between $V_{\rm max}/\sigma_p(0)$ and $r_{\rm max}/R_c$. The three 
thick lines (almost indistinguishable from one another in this panel) 
correspond to three different values of the tidal-to-core radius ratio 
chosen for the King model, and confirm the result anticipated above 
regarding the weak dependence on $R_t$ of these results. 
 
{\it Any} NFW halo whose circular velocity peaks somewhere along the 
thick curve in Figure~\ref{fig:s0rcrs} is, therefore, consistent with 
a King model of given $\sigma_p(0)$ and $R_c$. A few examples of NFW 
halos belonging to this ``family'' are shown by the thin lines in 
Figure~\ref{fig:s0rcrs}. Note that essentially all the circular 
velocity profiles of these halos, despite having very different masses 
and concentrations, cross each other at $r\approx R_c$. This implies 
that the total enclosed mass within the core radius of a King model is 
robustly determined given our assumptions: in particular, we find 
$V_c(R_c)\simeq 1.2\, \sigma_p(0)$ and $M(R_c)\simeq 1.44 \, R_c \, 
\sigma_p(0)^2/G$.  Physically, this means that, although the {\it 
total} mass and radial extent of the halo are not well pinned down, 
the mass within the luminous region sampled by the tracers is. (See 
also Strigari et al. 2007 for a similar approach and conclusion.) 
 
A further constraint may be gleaned from Fig.~\ref{fig:sprof}, which 
shows that the {\it shape} of the stellar velocity dispersion profile 
also depends on the degree of segregation between stars and dark 
matter. For $R_c \simless 0.1 \, r_s$ ($=0.05\, r_{\rm max}$) the 
velocity profile remains approximately flat well outside the core 
radius, and declines abruptly only at the ``tidal'' radius. On the 
other hand, less segregated King models show a steep velocity decline 
noticeable near the core radius: for example, for $R_c \sim r_s$ the 
velocity dispersion declines by roughly $70\%$ at the core radius from 
the central value. The velocity dispersion profiles of all dwarfs for 
which such data are available show little sign of declining outside 
$R_c$ (see references in \S~\ref{ssec:obs}). In the context of our 
modeling, this suggests that the stellar component is deeply embedded 
within its parent CDM halo (i.e., $R_c \simless \, r_{\rm max}$), an 
issue to which we return below. 
 
\subsection{Application to Milky Way dwarfs} \label{sec:MWhalos} 
 
One way of breaking the degeneracy illustrated in 
Figure~\ref{fig:s0rcrs} is to appeal to the results of cosmological 
N-body simulations. These show that there is a strong correlation 
between the mass and concentration of a CDM halo or, equivalently, 
between $r_{\rm max}$ and $V_{\rm max}$. As discussed by NFW, this 
correlation arises because the characteristic density of a halo is 
proportional to the density of the Universe at the time of its 
assembly. In practice, and for galaxy-sized systems, the mass-density 
dependence is quite weak, implying that $r_{\rm max}$ is roughly 
proportional to $V_{\rm max}$. This correlation is now well 
established, and a number of authors provide simple formulae to 
compute it once the cosmological parameters are specified (NFW, Eke, 
Navarro \& Steinmetz 2001, Bullock et al 2001). 
 
\begin{figure} 
\plotone{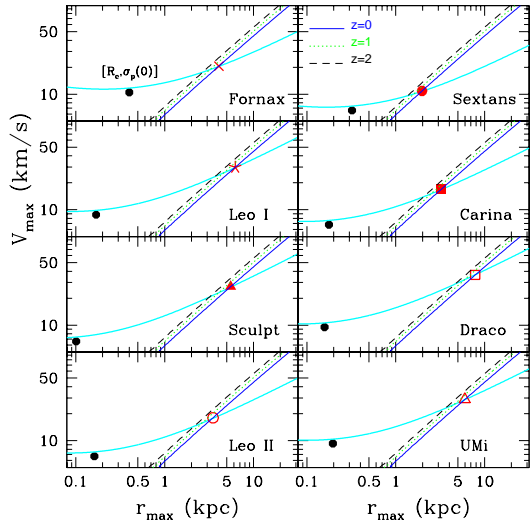} 
\caption{Each panel shows, for the 8 Milky Way dwarfs in our sample, 
the King-NFW degeneracy (curved line), as well as the predictions for 
$\Lambda$CDM cosmogony (set of straight lines). Only NFW halos at the 
intersection of both sets of curves (marked by a red symbol) are 
consistent with both cosmological constraints and the structure and 
kinematics of the dwarfs. The three sets of cosmological curves 
correspond to NFW halos identified at various redshifts; we adopt the 
$z=0$ models here, but note that our conclusions are unlikely to be 
severely affected by this choice.} 
\label{fig:vmrs_8} 
\end{figure}

The relation between $V_{\rm max}$ and $r_{\rm max}$ implies that, of 
all NFW halos in the ``family'' of models allowed for a given dwarf by 
the degeneracy illustrated in Figure~\ref{fig:s0rcrs}, a single one 
will be consistent with the parameters expected in a given 
cosmogony. This is shown in Figure~\ref{fig:vmrs_8}, where the curved 
lines show, in different panels and for each of the eight MW dSphs, the 
King-NFW ``degeneracy'' relation. The straight lines in each panel 
delineate the $V_{\rm max}$-$r_{\rm max}$ relation consistent with the 
$\Lambda$CDM cosmogony. The set of three straight lines correspond to NFW 
halos identified at $z=0$, $1$, and $2$, respectively. We shall 
hereafter ignore the relatively small difference between these curves 
and refer, for simplicity, all of our results to $z=0$. 
 
In each panel a solid dot indicates, for reference, the core radius 
and the central velocity dispersion of each galaxy, as listed in 
Table~\ref{tab:obs}. All of these points lie well to the left of the 
cosmological relations, confirming our earlier suggestion that the 
luminous components are significantly segregated within their dark 
halos (i.e., they are substantially smaller for given characteristic 
velocity). 
 
\begin{figure} 
\plotone{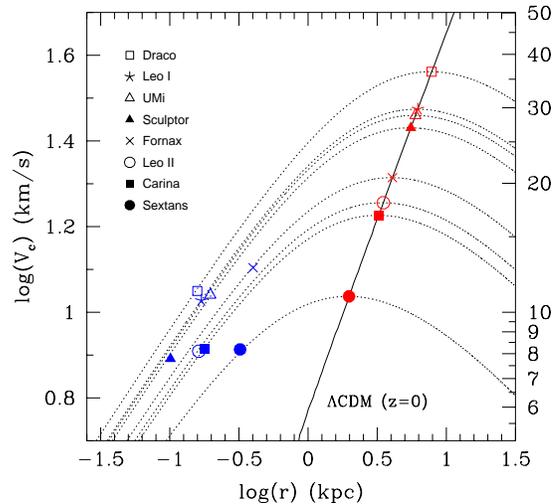} 
\caption{Circular velocity profiles of NFW halo models consistent with 
the observed structure and kinematics of the stars and with the 
$\Lambda$CDM cosmological constraints.  Blue symbols (on the left) 
denote the circular velocity at the core radius of each dwarf, where 
it is best constrained. Labels rank, from top to bottom, all dwarfs in 
decreasing order of halo mass.} 
\label{fig:vc} 
\end{figure}

The intersection between the King-NFW degeneracy and the cosmological 
relation is marked by red symbols for each dwarf. This indicates the 
parameters of the NFW halo model that is consistent with the 
$\Lambda$CDM cosmogony and, at the same time, matches the kinematics 
and structure of each dwarf galaxy.  The circular velocity profiles of 
the eight NFW halos satisfying these criteria are shown in 
Fig.~\ref{fig:vc}, and labelled from top to bottom in order of 
decreasing halo mass.  
 
This ranking shows interesting peculiarities. For example, it shows 
that the peak circular velocities of dwarf halos vary from $\sim 17$ 
to $\sim 35$ km/s, corresponding to a spread of about $8$ in mass, 
much narrower than the factor of $\sim 70$ spanned by dSph 
luminosities. Note as well that, as anticipated in \S1, halo mass is 
not monotonically related to luminosity. Intriguingly, Draco, one of 
the faintest dwarfs in our sample, is assigned the most massive halo, 
whereas Fornax, despite being 70 times brighter, is assigned a halo 5 
times less massive. 
 
\begin{figure} 
\plotone{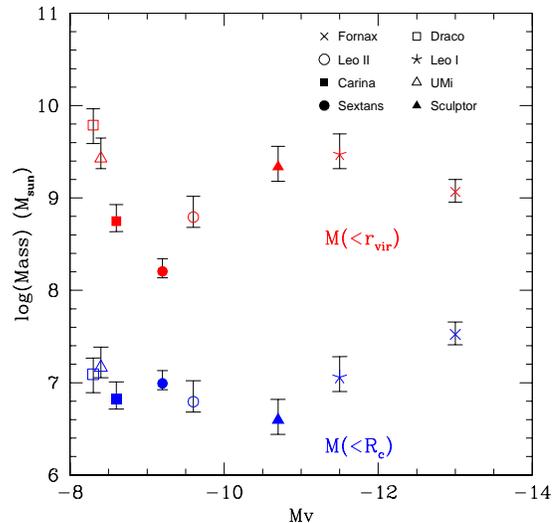} 
\caption{ Virial masses inferred from the NFW halo models shown in 
Fig.~\ref{fig:vc} for each of the 8 MW dwarfs. Also shown is the mass 
enclosed within $R_c$ for these systems. Error bars show the 
dependence of $M_{\rm vir}$ on the redshift adopted for the 
cosmological $V_{\rm max}=V_{\rm max}(r_{\rm max},z)$ relation. Top of 
the error bar corresponds to $z=0$, the bottom to $z=2$. Note that all 
dwarfs, irrespective of luminosity, have approximately the same mass 
within their core radii.} 
\label{fig:mvir} 
\end{figure}

The lack of correlation between luminosity and halo mass is shown 
explicitly in Figure~\ref{fig:mvir}, where we plot, as a function of 
total visual magnitude, the virial mass of the halo (red symbols at 
the top), as well as the mass within the core radius of each dwarf 
(blue symbols at bottom). Symbols are the same as those used in 
Fig.~\ref{fig:vc} to denote different galaxies.

Note that the total mass within the core radius---a fairly robust 
measure according to our discussion of Fig.~\ref{fig:s0rcrs}---is 
approximately independent of luminosity{\footnote{This is not a new 
result, and is consistent with Mateo's (1998) conclusion that simple 
dynamical mass estimates of dwarf galaxies are independent of 
luminosity.}}. This implies that the {\it average density} of dark 
matter will be higher in physically smaller systems such as Draco than 
in more extended dwarfs such as Fornax. Why does this matter? Because 
more massive halos are denser than less massive ones {\it at all 
radii} in the CDM cosmogony. Indeed, note that the circular velocity 
profiles of the NFW halos shown in Figure~\ref{fig:vc} do not cross, 
which means that measuring the halo circular velocity (or mass) at 
{\it any} radius leads to a well-defined estimate of the {\it total} 
mass of the halo. Because Draco, despite being faint, has a circular 
velocity comparable to Fornax's at a much smaller radius, it requires 
a denser, and therefore more massive, halo to satisfy the 
observational constraints. 
 
Of the eight dwarfs, the lowest halo mass corresponds to Sextans, 
whose relatively large radius and small velocity dispersion is 
inconsistent with a very massive CDM halo.

%\begin{figure} 
%\plotone{rho_all.ps} 
%\caption{ NFW averaged density profiles ($\langle \rho \rangle\equiv 
%V_c^2/r^2$) for the models shown in Fig.~\ref{fig:vc}. } 
%\label{fig:rho} 
%\end{figure} 
 
\subsection{Effects of tidal stripping} 
 
The results obtained in the previous section assume that the structure 
of dark matter halos is well approximated by an NFW profile. Although 
this assumption may be appropriate for isolated halos, it is unlikely 
to hold in detail for ``substructure'' halos of dSphs orbiting within 
the main halo of the Milky Way.  Recently, Stoehr et al (2002), 
Hayashi et al (2003) and Kazantzidis et al (2004) have examined the 
modifications undergone by an NFW halo as it is tidally stripped 
inside a more massive system. Stripping affects principally the outer 
regions of the halo, and therefore deeply embedded stellar structures 
may survive unscathed the removal of large fractions of their halos.

\begin{figure} 
\plotone{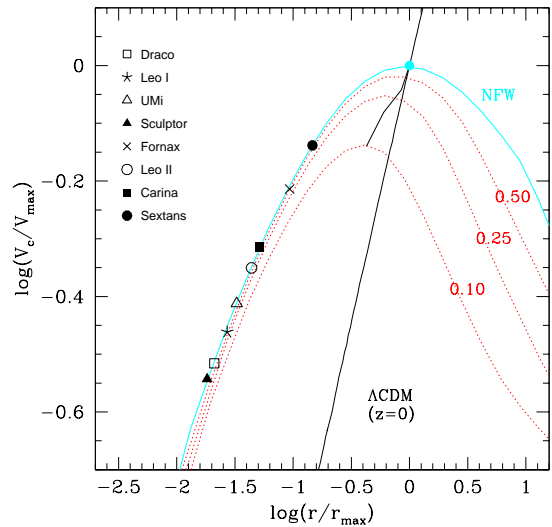} 
\caption{ NFW halo circular velocity, normalized to the location of 
the peak. The core radius of each of the MW dwarf galaxies is 
shown. Note that, except for Sextans, in all cases the core radius lies 
within $0.1\, r_{\rm max}$.  Dotted lines show the changes imposed by 
tidal stripping, as the original NFW halo loses $50\%$, $75\%$, and 
$90\%$ of its mass. Note that most dwarfs are sufficiently embedded 
within their halos that they would only be affected if the halo lost 
more than $\sim 90\%$ of their original mass.} 
\label{fig:strip} 
\end{figure}

This is illustrated in Figure~\ref{fig:strip}, which shows the 
circular velocity profile of an NFW halo, after being tidally stripped 
of 50\%, 75\%, and 90\% of its mass, respectively. The profiles are 
taken from N-body simulations of NFW halos orbiting within the 
potential of a much larger system, as in Hayashi et al (2003). 
Figure~\ref{fig:strip} indicates that NFW halos must lose at least 
90\% of their original mass before regions within $r \simless 0.1\, 
r_{\rm max}$ are significantly affected. Interestingly, this is 
precisely the region that our analysis suggests is populated by the 
stars of MW dwarfs. According to Fig.~\ref{fig:vc} the average 
$R_c/r_{\rm max}$ for all MW dwarfs is $0.054$ and they all have, with 
the possible exception of Sextans, $R_c \simless 0.1 \, r_{\rm max}$. 
 
Tidal stripping is clearly affecting the structure of the Sagittarius 
dSph (which is thus not in the sample considered here), but the 
evidence is less clear-cut in the case of other dwarfs, despite a rich 
literature on the topic. Much of this work is based on detecting stars 
associated with a dwarf beyond the King ``tidal radius'', or on the 
interpretation of ``breaks'' in the surface density profile in the 
outer regions of a dwarf (e.g. Ursa Minor, Mart\'{\i}nez-Delgado et 
al. 2001, Carina, Majewski et al. 2005).   
 
The presence of a break in the outer profile, however, does {\it not} 
imply that the system is necessarily losing stars. Such feature might 
just be intrinsic to the dSph, as shown, for example, by the 
multiple-component models presented in McConnachie et 
al. (2006). These authors show that systems with multiple stellar 
components (such as Andromeda II) may have overall density profiles 
that resemble those of dwarf galaxies with outer ``tidal'' breaks. The 
presence of well-defined and kinematically-distinct populations of 
stars within some dSphs (Sculptor, Tolstoy et al. 2004;  Fornax, Battaglia et al. 2006; Canis Venatici, Ibata et al. 2006), indeed, 
argues {\it against} tides as a major driver in the evolution of a 
dSph, since tidal stirring would tend to uniformize such distinctions. 
 
On similar grounds, one cannot rule out that other features in the 
light profile usually ascribed to tides, such as the presence of lumps 
(UMi, Olszewski \& Aaronson 1985), shells (Fornax, Coleman et 
al. 2005) and aspherical isopleths (UMi, Mart\'inez-Delgado 2001) may 
actually reflect complexities in the formation process and/or the 
internal evolution of dwarf galaxies. 
 
The lack of overwhelming evidence for ongoing stripping of stars from 
most dSphs may thus be taken to imply that dwarf halos have retained 
at least 10\% of their original mass at present. This is important, 
because even for such sizeable loss, the peak circular velocity of the 
halo is barely affected. 
 
The thin line in Figure~\ref{fig:strip} tracks the position of the 
circular velocity peak of an NFW halo as it is tidally stripped, and 
shows that halos that have lost 75\% of their mass to stripping see 
only a 10\% decrease in $V_{\rm max}$. Even after losing 90\% of its 
mass, a halo sees its $V_{\rm max}$ reduced by only $\sim 
30\%${\footnote{The {\it location} of the peak, $r_{\rm max}$, on the 
other hand, is more significantly affected, and shifts inwards by 
almost a factor of $\sim 3$ after a halo loses 90\% of its mass.}}. 
We conclude, therefore, that, unless dSphs have suffered catastrophic 
tidal losses, our estimates of the peak circular velocities for the MW 
dSphs (Fig.~\ref{fig:vc}) should be relatively robust, even if the 
amount of dark matter beyond the luminous radius remains unknown. 
 
We emphasize, however, that tides do pose a number of interesting 
questions that our modeling fails to address. For example, do King 
profiles survive strong tidal stripping?. Do we expect the presence of 
``extra-tidal'' stars beyond a {\it break radius} if the luminous 
component is still surrounded by dark matter? These issues are best 
addressed by self-consistent N-body simulation of the tidally-driven 
evolution of multi-component dSph models. This is beyond the scope of 
the present paper, but will be dealt with in a forthcoming paper of 
our series. 
 
\begin{figure} 
\plotone{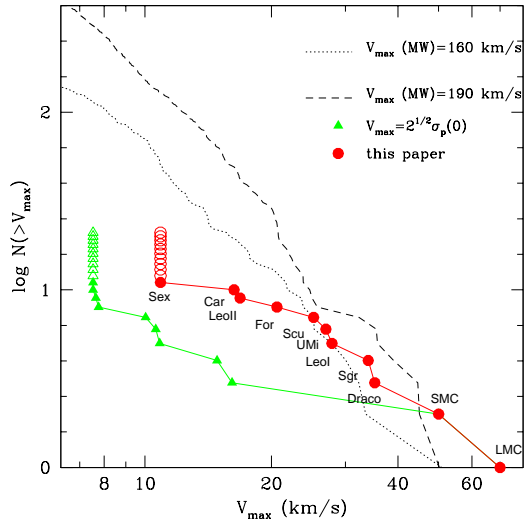} 
\caption{Cumulative number of substructures as a function of maximum 
circular velocity for two Milky Way-sized $\Lambda$CDM halos (solid 
and dashed lines, data from Diemand et al. 2006). Triangles 
approximate the maximum circular velocity of the Milky Way satellites 
as $V_{\rm max}\simeq \sqrt{2}\sigma_p(0)$ (Klypin et al. 1999, Moore 
et al. 1999) whereas circles show the estimates of $V_{\rm max}$ 
obtained in this contribution. For the LMC and SMC we assume that 
$V_{\rm max}$ equals the maximum rotation velocity of the HI disk (Kim 
et al 1998, Stanimirovi{\'c} et al 2004). Open symbols denote systems 
with measured and lacking velocity dispersion measurements, 
respectively. Note that our estimates considerably alleviate the 
``satellite crisis'' for satellites with peak circular velocities 
exceeding $10$-$20$ km/s.} 
\label{fig:vccum} 
\end{figure}

\begin{figure} 
\plotone{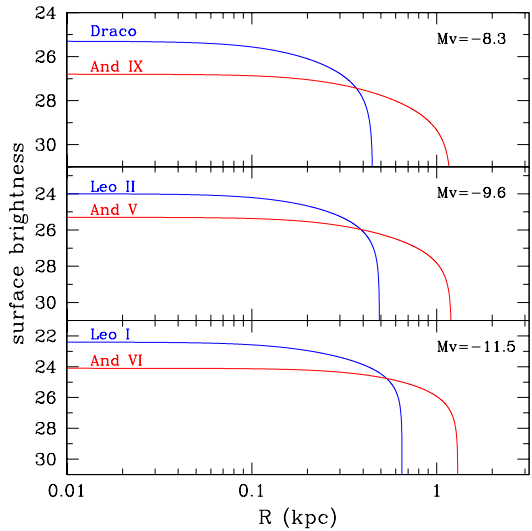} 
\caption{Comparison of the surface density profile of pairs of dwarf 
galaxies of similar luminosity, one orbiting around M31 and the other 
around the Milky Way. On average M31 dwarfs are larger and have lower 
surface brightness than their MW counterparts.} 
\label{fig:mw_m31} 
\end{figure}

\subsection{Application to the ``satellite crisis''} 
 
The analysis of the preceding subsections suggests that the 8 MW 
dwarfs chosen for our analysis inhabit halos with peak circular 
velocities in the range $17$-$35$ km/s, a factor of $\sim 3$ higher 
than their central (stellar) velocity dispersions.  As discussed by 
Stoehr et al (2002), Hayashi et al (2003) and Kazantzidis et al 
(2004), this correction is enough to alleviate substantially the 
``satellite crisis'' highlighted by Klypin et al (1999) and Moore et 
al (1999). This is illustrated in Fig.~\ref{fig:vccum}, which compares 
the estimates of the peak circular velocities obtained in this 
contribution (full circles) with those that adopt mass-follows-light 
models (full triangles). 
 
In rough terms, N-body simulations show that there are typically about 
$20$-$30$ substructure halos with peak circular velocities exceeding 
$10\%$ of the main halo's virial velocities. That is shown in 
Fig.~\ref{fig:vccum} from the results of Diemand, Kuhlen \& Madau 
(2006) for two different Milky Way-like halos (dotted and 
dahsed lines).  According to our analysis, there are about a dozen 
dwarfs with $V_{\rm max} \simgreat 17$ km/s (adding to our sample 
brighter satellites such as the Magellanic Clouds).  This velocity 
corresponds to between $8$ and $11\%$ of the MW virial velocity, 
assuming, as seems likely, that the latter is in the range $150$-$220$ 
km/s. This factor of $\sim 2$--$3$ discrepancy between the number of 
massive dark matter substructures and luminous satellites does not 
appear extravagant given the uncertainties.  
 
Furthermore, the discrepancy may disappear altogether if (as the 
scenario outlined in \S~\ref{sec:int} would suggest) the ultra-faint 
MW dwarfs identified in SDSS data (open symbols in 
Fig.~\ref{fig:vccum}) turn out to inhabit halos of masses comparable 
to those of the 8 dSphs we consider here. Since these newly discovered 
dwarfs have, on average, similar physical size to the dSphs in Table 
1, we expect that their velocity dispersions will also be comparable 
and in the range $\sim 6$-$10$ km/s, a prediction of this 
CDM-motivated modeling that should be testable in the near future. 
 
Indeed, further progress on this subject seems imminent, as 
observational campaigns secure velocity dispersions for many of the 
newly discovered dwarfs and extend available data to allow for 
velocity dispersion profiles to be measured for more systems (see, 
e.g., the recent preprint by Simon \& Geha 2007). One also expects 
that numerical simulations will improve to the point that a reliable 
direct estimate of the number of substructure halos satisfying the 
$M(R_c)$ constraints shown in Fig.~\ref{fig:mvir} will be possible, 
obviating the need to extrapolate results out to the peak of the halo 
circular velocity curve. The latter goal requires simulations able to 
resolve convincingly the inner $\sim 100$ pc of substructure halos, an 
order of magnitude improvement over simulations published so far. This 
is a numerical challenge that will likely be met soon but that will 
nevertheless require the investment of massive computational resources 
(see the recent preprint by Kuhlen et al 2007 for a report of recent 
progress on this issue.) 
 
\begin{figure} 
\plotone{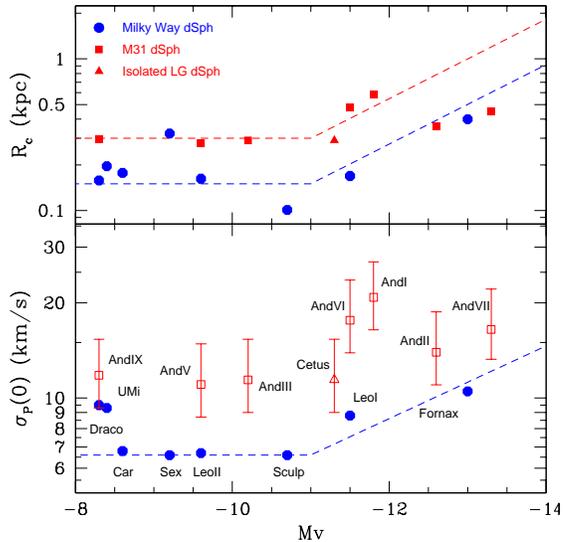} 
\caption{{\it Upper panel}: Core radii of Milky Way (filled circles) 
and M31 (filled squares) dwarf galaxies as a function of their 
absolute magnitude. The dashed lines are intended to guide the eye and 
to highlight the difference in size between the M31 and MW dwarf 
populations.  {\it Bottom panel}: Measured (filled symbols) and 
{\it predicted} (open symbols) central velocity dispersions if the Milky Way 
and M31 dSph's are embedded in similar dark matter halos. See text for details.} 
\label{fig:s0pred} 
\end{figure} 
 
\begin{figure} 
\plotone{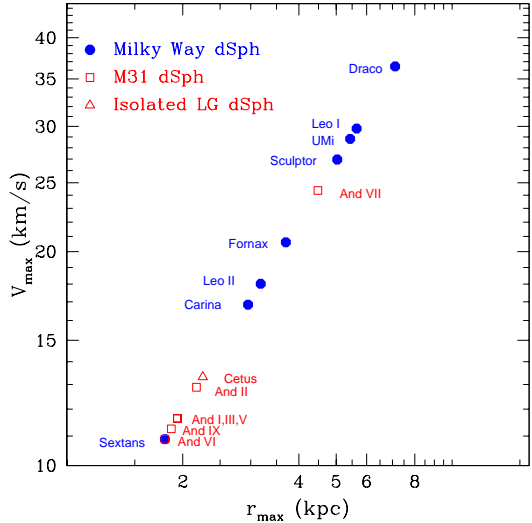} 
\caption{NFW halo parameters $(r_{\rm max}, V_{\rm max})$ for the 
Milky Way (filled circles) and M31 (open squares) dwarf galaxies 
expected if the velocity dispersion of M31 dwarfs are found to be 
similar to those of their MW counterparts. Note that these assumptions 
lead to dark matter subhalos in M31 that are substantially smaller than 
those in the Milky Way, a result which is disfavored by the present 
analysis. See text for details.} 
\label{fig:rsvmpred} 
\end{figure} 
 
\section{Milky Way vs. M31 dwarfs}\label{sec:m31} 
 
Photometric studies of the dwarf galaxies in the Local Group (Irwin \& 
Hatzidimitriou 1995, McConnachie \& Irwin 2006) have shown that, at 
fixed luminosity, M31 dwarfs are considerably more extended than their 
Milky Way counterparts (see Table~\ref{tab:obs}).  We illustrate this 
in Fig.~\ref{fig:mw_m31}, where King model fits to the surface 
brightness profiles of three pairs of dwarfs of similar luminosity are 
shown. Each pair consists of one M31 satellite and one MW satellite, 
respectively. Clearly, the M31 dwarfs are systematically different: 
they are about 1.5 mag fainter in central surface brightness and about 
a factor of two larger in radial extent compared with MW dwarfs (see 
top panel of Fig.~\ref{fig:s0pred}). 
 
The origin of this difference is unclear, but it offers nonetheless a 
way of assessing the general validity of the modeling proposed in the 
previous sections. Indeed, under the plausible assumption that the 
dark matter halos of dwarfs in M31 and in the Milky Way are not 
systematically dissimilar, the arguments of \S~\ref{ssec:seg} imply 
that the difference in size of M31 dwarfs should be reflected in their 
kinematics through significantly higher velocity dispersions. Note 
that this prediction depends explicitly on the presence of dark matter 
halos whose mass and extent are largely independent from the stars', 
as implicitly assumed in our modeling so far. Other (also plausible?) 
assumptions, such as mass-traces-light, would lead to the {\it 
opposite} trend, and would predict systematically {\it lower} velocity 
dispersions. 
 
In order to estimate quantitatively the velocity dispersion of M31 
dwarfs we need to know how spatially segregated the stars are from the 
dark matter in these systems. Once this is fixed by, for example, 
assuming a value for the ratio $R_c/r_{\rm max}$, we can compute 
$r_{\rm max}$ for each dwarf and, from that, infer $V_{\rm max}$ 
assuming the cosmological relation between these parameters. Since the 
ratio $R_c/r_{\rm max}$ also fixes $V_{\rm max}/\sigma_p(0)$ according 
to the King-NFW degeneracy shown in Fig.~\ref{fig:s0rcrs}, an estimate 
can then be made of the central velocity dispersion, $\sigma_p(0)$, 
for each M31 dwarf. 
 
We have followed this procedure, assuming $R_c/r_{\rm max}=0.05$, 
consistent with the average segregation measure derived for MW dwarfs 
(see Fig.~\ref{fig:strip}), as well as with the flat $\sigma_p(R)$ 
constraint mentioned in \S~\ref{ssec:seg}. This leads to the estimates 
of the velocity dispersions of M31 dwarfs plotted in the bottom panel 
of Fig.~\ref{fig:s0pred} (open symbols). Error bars show the variation 
induced in the estimate by allowing $R_c/r_{\rm max}$ to vary between 
$0.025$ and $0.1$. Clearly, the velocity dispersion of M31 dSphs is 
significantly larger (by about a factor of $\sim 2$) than that of 
their MW counterparts, as shown by the shift between solid circles and 
open symbols in Fig.~\ref{fig:s0pred}. 
 
We emphasize that these ``predictions'' for the velocity dispersions 
of M31 dwarfs should be taken with caution: indeed, the open symbols 
in Fig.~\ref{fig:s0pred} are best interpreted as {\it indicative} of 
the relative shift in the velocity dispersion of the {\it population} 
of M31 dwarfs rather than as separate individual predictions.  
 
So far, the few available observations are consistent with the 
predicted trend. Of the three dwarfs outside the MW for which velocity 
dispersions have been published (see, Table~\ref{tab:obs}), the most 
reliable is Cetus and, at $17$ km/s, has the highest $\sigma_p$ in the 
sample, consistent with the predicted trend (our naive prediction 
yields $11.4^{+4}_{-3}$ km/s). Likewise, the higher of the two estimates derived 
by Chapman et al (2005, $12$ km/s) for And IX is also consistent with 
this trend, although the very few stars in this galaxy (as well as in 
And II) with measured velocities suggest that it would be premature to 
draw firm conclusions until the data improves substantially. 
 
It is clear, however, that should future observations fail to confirm 
the predicted trend, one or many of our assumptions would need to be 
revised. As an illustrative example, we consider how our 
interpretation would change if the velocity dispersion of M31 dwarfs 
were to show no systematic departure from that of their MW 
counterparts (i.e., if the open symbols were found to align with the 
dashed line in the bottom panel of Fig.~\ref{fig:s0pred}). In this 
case, one would be forced to conclude that M31 and MW dwarfs of 
similar luminosity inhabit systematically different halos. In 
particular, M31 dwarf halo hosts would need to be significantly less 
massive (i.e., lower $V_{\rm max}$ and smaller $r_{\rm max}$) in order 
to accommodate their larger physical size but similar kinematics. 
 
This is shown in Fig.~\ref{fig:rsvmpred}, and imply that essentially 
all M31 dwarfs considered here would have $V_{\rm max} \simless 15$ 
km/s (with the possible exception of And VII). Since, in all 
likelihood, M31 inhabits a halo at least as massive as the Milky 
Way's, it should have at least as many massive substructures as our 
own Galaxy, and one would need to explain why those substructures have 
not ``lit up'' in M31 as they have in the Milky Way. We consider this a 
compelling argument against this interpretation and in support of 
higher velocity dispersions for M31 dwarfs as the most natural 
prediction of CDM-motivated models of the Local Group dSphs. 
 
One last possibility should be mentioned; namely, that the structural 
differences between M31 and MW dwarfs reflect differences in the 
dynamical evolution driven by tides in M31 and the Galaxy. Although we 
have argued in \S 3.3 that this is unlikely, a definitive assessment 
requires high-resolution, self-consistent N-body simulations of how 
tides affect multiple-component models of dSph galaxies. We plan to 
address this issue in the follwoing paper of this series (Pe\~{n}arrubia, Navarro \& 
McConnachie, in preparation).

%\begin{figure} 
%\plotone{m31.comp.ps} 
%\caption{{\it Upper panel}: Predicted velocity dispersion profile of 
%And V assuming that both it and Leo II are embedded in identical dark 
%matter halos and $R_c/r_s=0.1$. Arrows denote the values of the core 
%radii for each system. {\it Bottom panel}: Predicted circular velocity 
%curve for And V if its central velocity dispersion is found to be the 
%same as Leo II. Note that in this last case the putative halo of the 
%M31 dwarf galaxy would have a larger scale radius $r_{s,{\rm M31}}=2 
%r_{s,{\rm MW}}$ and the same maximum velocity. } 
%\label{fig:m31comp} 
%\end{figure} 

\section{Summary}\label{sec:disc} 
 
We have considered in this paper the observational constraints placed 
on the mass and spatial extent of cold dark matter halos surrounding 
Local Group dwarf spheroidal galaxies. Assuming that the luminous 
component may be approximated by King models and that dark halos 
follow the NFW mass profile, we conclude that estimates of the halo 
mass of a dSph depend principally on the degree of spatial segregation 
between stars and dark matter: the more embedded the stars within the 
dark halo the more massive the halo must be in order to explain the 
observed stellar kinematics. This degeneracy may be broken by 
appealing to the results of cosmological N-body simulations---which 
indicate a strong correlation between mass and size of cold dark 
matter halos. The procedure results in reasonably robust estimates of 
the dark halo mass of a dSph and of its peak circular 
velocity. 
 
Our analysis indicates that the mass within the luminous radius of the 
dwarf is well constrained by the velocity dispersion of the stars, and 
that it is approximately independent of the luminosity of the 
system. Within the context of cosmologically motivated CDM halos, this 
implies that dSphs are surrounded by halos with peak circular 
velocities a factor of $\sim 3$ times larger than their stellar 
velocity dispersion. These results are consistent with previous work, 
and substantially alleviate the ``missing satellites'' problem, 
particularly if most of the newly discovered ultra-faint satellites 
are confirmed to be bona-fide dwarf galaxies akin to the more 
luminous, classical systems analyzed here. 
 
We also find that stars in dSphs are deeply embedded within their dark 
matter halos, with core radii typically of order 5\% the radius where 
the original halo circular velocity peaks. This yields velocity 
dispersion profiles that are nearly flat out to the nominal King tidal 
radius, as observed for most of the Milky Way dSphs with suitable 
kinematical data. The deep segregation of the sellar component also 
implies that the stars of dwarf galaxies are fairly resilient to tidal 
disruption; a halo would need to lose more than 90\% of its mass 
before stars begin to be affected.  Even in this case, the peak 
circular velocity of a halo would only drop by $\sim 30\%$, suggesting 
that our $V_{\rm max}$ estimates are robust even if halos have been 
heavily (but perhaps not catastrophically) affected by tidal 
stripping. 
 
Applied to the dSph population of M31, this analysis suggests that the 
systematic difference in size between M31 dwarfs and their MW 
counterparts should be reflected in their kinematics. Under the 
plausible assumption that substructure halos in M31 are similar to 
those of the Milky Way, we conclude that, as a population, the 
velocity dispersion of M31 dwarfs should be $\sim 50$--100\% higher than 
that of dwarfs in our own Galaxy.  
 
This prediction is likely to be validated (or challenged) soon by the 
results of the many ongoing observational projects devoted to 
obtaining accurate spectra and radial velocities for stars in these 
galaxies. Whether Local Group dwarfs conform with the expectations of 
the prevailing CDM paradigm or throw it a further gauntlet we should 
know in the near future. 
 
\acknowledgments 
 
This work has been supported by various grants to JFN from the Natural 
Sciences and Engineering Research Council of Canada (NSERC). JP thanks 
Scott Chapman for all his help. We are very grateful to 
D. Mart\'inez-Delgado and B. Gibson for useful comments. We also thank 
J. Diemand for kindly providing the data plotted in 
Fig.~\ref{fig:vccum}.  JFN acknowledges useful discussions with Simon 
White and thanks Carlos Frenk and the Institute of Computational 
Cosmology of the University of Durham for their hopsitality during the 
time that this work was carried out. 
 
{} 
  
\end{document}